\documentclass{aa}
\usepackage{graphicx}
\def\Msolar{\hbox{${\rm M}_\odot$}}

\begin{document} 

\title{Why is the mass function of NGC\,6218 flat?\thanks{Based on
observations collected at the European Southern Observatory, Paranal,
Chile, as part of programme 63.L-0423.}}

\author{Guido De Marchi \inst{1}, Luigi Pulone \inst{2} and Francesco
Paresce \inst{3}
}

\offprints{G. De Marchi}

\institute{ESA, Space Science Department, Keplerlaan
1, 2200 AG Noordwijk, Netherlands, gdemarchi@rssd.esa.int 
\and
INAF, Osservatorio Astronomico di Roma, Via di Frascati 33, 00040 Monte
Porzio Catone, Italy, pulone@mporzio.astro.it
\and
INAF, Viale del Parco Mellini 84, 00136 Roma, Italy, fparesce@inaf.it}

\date{Received 6 July 2005; Accepted 28 November 2005}
\titlerunning{The mass function of NGC\,6218}
\authorrunning{De Marchi et al.}

\abstract{ 

We have used the FORS-1 camera on the VLT to study the main sequence
(MS) of the globular cluster NGC 6218 in the $V$ and $R$ bands. The
observations cover an area of $3\farcm4 \times 3\farcm4$ around the
cluster centre and probe the stellar population out to the cluster's
half-mass radius ($r_{\rm h} \simeq 2\farcm2$). The colour-magnitude
diagram (CMD) that we derive in this way reveals a narrow and well
defined MS extending down to the $5\, \sigma$ detection limit at $V
\simeq 25$, or about $6$ magnitudes below the turn-off, corresponding
to stars of $\sim 0.25$\,\Msolar. The luminosity function (LF) obtained
with these data shows a marked radial gradient, in that the ratio of
lower- and higher-mass stars increases monotonically with radius. The
mass function (MF) measured at the half-mass radius, and as such
representative of the cluster's global properties, is surprisingly
flat. Over the range $0.4 - 0.8$\,\Msolar, the number of stars per unit
mass follows a power-law distribution of the type $dN/dm \propto
m^{0}$, where, for comparison, Salpeter's IMF would be $dN/dm \propto
m^{-2.35}$. We expect that such a flat MF does not represent the
cluster's IMF but is the result of severe tidal stripping of the stars
from the cluster due to its interaction with the Galaxy's gravitational
field. Our results cannot be reconciled with the predictions of recent
theoretical models that imply a relatively insignificant loss of stars
from NGC\,6218 as measured by its expected very long time to
disruption. They are more  consistent with the orbital parameters based
on the Hipparcos reference system that imply a much higher degree of
interaction of this cluster with the Galaxy than assumed by those
models. Our results indicate that, if the orbit of a cluster is known,
the slope of its MF could be useful in discriminating between the
various models of the Galactic potential.

}
\maketitle

\section{Introduction}

A satisfactory understanding of the properties of the initial mass
function (IMF) of globular clusters (GCs) is a major objective of
current astrophysical research in that GCs are the closest example of
star formation at high redshift (Krauss \& Chaboyer 2003). The most
reliable observations so far available that reach near the bottom of
the stellar MS indicate that all halo clusters have a very similar
present global MF (PGMF), which peaks at $\sim 0.35$\,\Msolar (Paresce
\& De Marchi 2000; De Marchi, Paresce \& Portegies Zwart 2005). The
internal dynamical relaxation process via stellar encounters is now
reasonably well understood (Meylan \& Heggie 1997) and validated 
observationally (e.g. De Marchi, Pulone \& Paresce 2000; Albrow, De
Marchi \& Sahu 2002; Pasquali et al. 2004) and allows us to derive the
global properties of the MF from a limited number of measurements
within a cluster.

Nevertheless, any hope to infer useful information on the properties of
the IMF from the observed present global MF (PGMF) rests on our ability
to roll back the effects that the tidal field of the Galaxy has exerted
on the stellar population of the clusters (Paresce \& De Marchi 2000).
Gravitational shocking due to repeated  interactions with the bulge and
disc of the Galaxy profoundly disrupt the original mass distribution by
ejecting low mass stars from the core and by compressing the tidal
boundary in phase space at each encounter. These phenomena, integrated
over the orbit and time and eventually causing the disruption of the 
cluster, can substantially alter the shape of the MF thereby completely
masking the properties of the IMF (Vesperini \& Heggie 1997). 

Theoretical models describing the interaction of GCs with the Galactic
tidal field have become progressively more detailed and, possibly,
accurate in the past decade or so. This has been made possible by an
in-depth analysis of the mechanisms responsible for cluster disruption
(Aguilar, Hut \& Ostriker 1988; Gnedin \& Ostriker 1997) and by the
availability of more accurate space motion parameters for the clusters
(Dauphole et al. 1996; Odenkirchen et al. 1997). Models that make use
of GC proper motion information (Dinescu et al. 1999; Baumgardt \&
Makino 2003) should, in principle, provide a more reliable description
of the clusters' past dynamical history than those purely based on
radial velocity data (Gnedin \& Ostriker 1997). On the other hand, the
models of Gnedin \& Ostriker (1997) explain rather convincingly, albeit
in a statistical sense, why the GCs that we see today are not randomly
distributed in parameter space but rather occupy regions of low
probability of disruption.  

The most useful indicator of the past dynamical history of GCs that
models of this type produce is the time to disruption, $T_{\rm d}$,
namely the time over which a cluster would be completely dissolved by
tidal forces. Unfortunately, this parameter is not directly observable,
thereby making it more difficult to assess the validity of the
models. It is, therefore, necessary to relate $T_{\rm d}$ to other
measurable parameters.

A rather obvious indication of a cluster's tidal disruption would seem
to be the presence of an extended tidal tail (Leon, Meylan \& Combes
2000; Odenkirchen et al. 2003). The unequivocal detection of tidal
tails, however, is hard to achieve on the basis of photometric
information alone (Baumgardt \& Kroupa 2005), which is often the only
available data. A more robust, and potentially more powerful approach
consists in looking at the properties of the MF of MS stars and at its
variations within the cluster, with respect to that of other well
behaved reference GCs.  This has allowed us to identify, for the first
time, the clear signature of tidal disruption in the cluster,
NGC\,6712, in the form of a severe  depletion of low-mass stars (De
Marchi et al. 1999; Andreuzzi et al. 2001). 

While our discovery has proved that GCs are indeed subject to the
effects of tidal stripping, the widely different values of $T_{\rm d}$
predicted by various models for NGC\,6712 have, at the same time,
raised  concerns as to the validity of the present physical
understanding of the processes involved. In order to better understand
the possible magnitude of the underlying discrepancy, we have studied
the properties of the MF of another GC, NGC\,6218, which, according to
all three sets of presently available models (Gnedin \& Ostriker 1997;
Dinescu et al. 1999; Baumgardt \& Makino 2003) should have experienced
an insignificant or very mild interaction with the Galactic tidal
field.  We would expect that its PGMF should accurately reflect the IMF
and our  aim was to use NGC\,6218 as a reference for NGC\,6712 and
other clusters.  In this paper we report on the properties of its MF
that shows that there may be something terribly wrong with these models
or, more likely, with their assumptions. 

The structure of the paper is as follows. The observations and their
reduction are described in Section\,2, whereas the results of the
photometry are discussed in Section\,3. Section\,4 is devoted to the LF
and MF of MS stars at various locations in the cluster. The dynamical
structure of NGC\,6218, as derived from these data, is presented in
Section\,5 and the overall implications for the understanding of the
interaction between GCs and the potential field of the Galaxy are
discussed in Section\,6.

\section{Observations and data analysis}

The observations on which this paper is based were obtained in 1999
June with the FORS1 camera at ESO's Very Large Telescope UT1, using the
high resolution collimator with a field of view (FOV) of
$3\farcm4$ on a side and a plate-scale of $0\farcs1$ per pixel. The log
of the observations, carried out in service observing mode, is given in
Table\,1. Seeing conditions were excellent on both observing nights (1
June and 16 June), and the point spread function (PSF) had an average
full width at half maximum of $0\farcs6$. Observations were carried out
in the Bessell $V$ and $R$ bands, with the exposure times listed in
Table\,1. The total equivalent  exposure time corresponds to 2160\,s in
V and 1440\,s in R.

The telescope was pointed at the nominal centre of NGC\,6218 at 
RA(J2000)$=16^{\rm h} 47^{\rm m} 14^{\rm s}5$ and DEC(J2000)$=-1^{\rm
o} 56\arcmin 52\farcs1$, corresponding to Galactic coordinates
$l=15.7$ and $b=26.3$. Thanks to the relatively wide FOV of FORS-1,
our observations reach out to the cluster's half-mass radius ($r_{\rm
h}=2\farcm16$; Harris 1996) and comfortably contain the region within
twice the core radius ($r_c=0\farcm72$; Harris 1996).

\begin{table}
\centering 
\caption{Log of the observations}
\begin{tabular}{cccc} \hline
 Date UT & Filter & Exp. time (s) & Air mass\\
\hline      
 1999 Jun 01 06:46:24 & V  & 180  & 1.232 \\ 
 1999 Jun 01 06:51:08 & V  & 180  & 1.246 \\ 
 1999 Jun 01 06:55:52 & V  & 180  & 1.261 \\ 
 1999 Jun 01 07:00:36 & V  & 180  & 1.277 \\ 
 1999 Jun 16 02:27:00 & R  & 180  & 1.163 \\
 1999 Jun 16 02:31:45 & R  & 180  & 1.154 \\
 1999 Jun 16 02:36:29 & R  & 180  & 1.146 \\ 
 1999 Jun 16 02:41:13 & R  & 180  & 1.138 \\	   
 1999 Jun 16 02:46:07 & V  & 180  & 1.131 \\ 
 1999 Jun 16 02:50:51 & V  & 180  & 1.124 \\ 
 1999 Jun 16 02:55:35 & V  & 180  & 1.118 \\ 
 1999 Jun 16 03:00:19 & V  & 180  & 1.112 \\ 
 1999 Jun 16 03:05:13 & R  & 180  & 1.107 \\
 1999 Jun 16 03:09:57 & R  & 180  & 1.103 \\
 1999 Jun 16 03:14:41 & R  & 180  & 1.099 \\ 
 1999 Jun 16 03:19:25 & R  & 180  & 1.095 \\
 1999 Jun 16 04:06:32 & V  & 180  & 1.086 \\ 
 1999 Jun 16 04:11:17 & V  & 180  & 1.087 \\ 
 1999 Jun 16 04:16:01 & V  & 180  & 1.089 \\ 
 1999 Jun 16 04:20:45 & V  & 180  & 1.092 \\ \hline
\end{tabular}
\vspace{0.5cm}
\label{tab1}
\end{table}

\begin{figure}
\centering
\resizebox{\hsize}{!}{\includegraphics{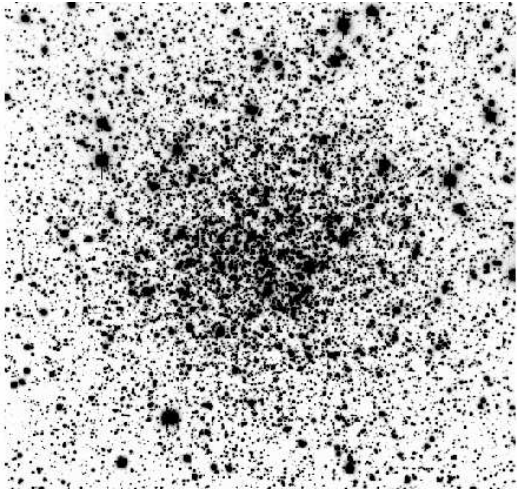}}
\caption{Negative image of the cluster NGC\,6218 of 180\,sec duration in
the Bessell $V$ band. North is up and East to the left. The field spans
$3\farcm4$ on a side.}
\label{fig1}
\end{figure}
 
The raw data were processed using the standard VLT pipeline calibration
and photometry was later performed using the DAOPHOTII/ALLFRAME package
(Stetson 1987; 1994). In particular, we first performed a photometric
reduction using the $\it daophot$, and $\it allstar$ tasks in order to
build a preliminary list of stars on each single exposure. These lists
were used to compute the coordinate transformation between each
individual frame and a reference image. All the  exposures, regardless
of their wavelength band, could thus be matched and combined to obtain
a median image, free from cosmic ray hits  and with the highest
signal-to-noise ratio (SNR) for the final star finding procedure. The
latter was performed in two steps, first by applying the routines {\em
daofind} and {\em allstar} on the stacked, deep  multi-filter image, by
setting the detection threshold at $5\,\sigma$ above the local
background level. We then analysed the PSF-subtracted image to recover
objects missed in the first step. The final star list was then used as
input to {\em allframe} for  simultaneous PSF-fitting photometry of all
of the objects in the individual frames. In order to take proper
account of the geometric distortion of the optics, we have devised an
automated procedure to define the best PSF candidate objects in each
region of our target field.  

The overall procedure above detected about $16,000$ bona-fide objects
in NGC\,6218. Their instrumental magnitudes were finally transformed to
the standard Johnson system by using the  photometric standard field
PG\,2213, observed with the same instrumental set-up during our runs. 

Our $5\,\sigma$ detection limits correspond to magnitudes $V\simeq 25$
and $R\simeq 24$, where the photometric error reaches $\sim 0.1$ mag. 

\begin{figure}
\centering
\resizebox{\hsize}{!}{\includegraphics[width=16cm]{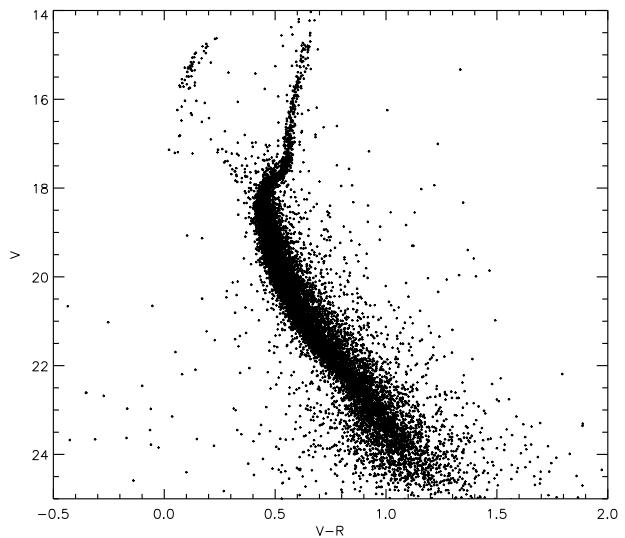}}
\caption{Colour--magnitude diagram of an area of $3\farcm4 \times
3\farcm4$ around the centre of NGC\,6218} 
\label{fig2}
\end{figure}

\section{The colour--magnitude diagram}

The $V$, $V-R$ CMD of the complete set of $16,015$ objects that we have
detected is  shown in Figure\,\ref{fig2}. At magnitudes brighter than
$V\simeq 18$ the CMD clearly shows the sequences of stars evolved off
the MS (sub-giant,  red-giant and horizontal branches), as well as a
conspicuous population of blue straggler stars. All these objects are
discussed in detail in a companion paper (Sabbi et al. 2005). Here we
concentrate on the cluster MS, which is narrow and well defined from
the turn-off at $V=18.8$, where the photometric error is small
($\sigma_{\rm V}=0.01, \sigma_{\rm R}=0.01$), through to $V=24$, where
the error on the magnitude grows to $\sigma_{\rm V}=0.04$ and
$\sigma_{\rm R}=0.04$. At $V\simeq 25$, where  our $5\,\sigma$
detection limit for the star-finding algorithm falls, the increasing
photometric error makes it difficult to distinguish the MS from
possible contaminating field stars. The photometric errors and
detection limit given here represent an average value over the whole
image, but given the considerable density gradient of our target field
the detection limit is brighter in the innermost regions and result in
a lower photometric completeness there, as we explain in Section\,4. 

\begin{figure}
\centering
\resizebox{\hsize}{!}{\includegraphics[width=16cm]{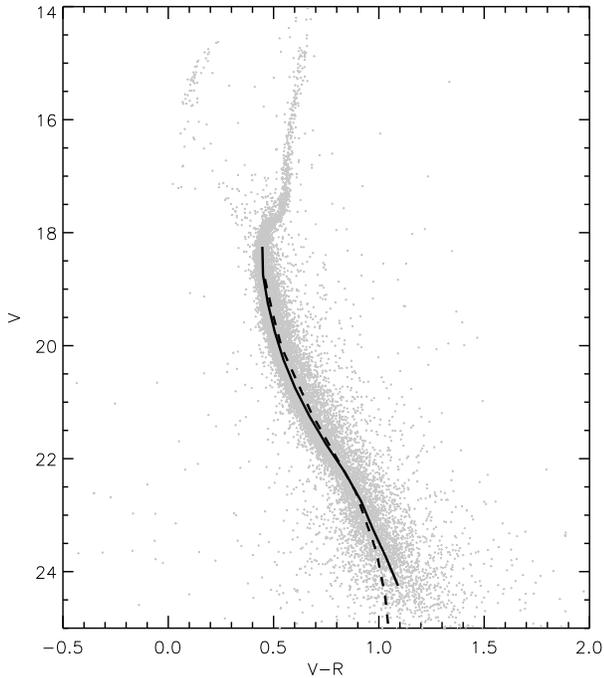}}
\caption{Main sequence ridge line of NGC\,6218 (solid curve) compared 
with the theoretical isochrones of Baraffe et al. (1997; dashed line)
for an age of 10\,Gyr and metallicity $[M/H]=-1.0$. The best fit of the
isochrones to the observed MS requires a distance modulus $(m-M)_{\rm
V}=14.23$ and a colour excess $E(B-V)=0.18$, consistent with the most
recent estimates of Hargis \& Sandquist (2004).}
\label{fig3}
\end{figure} 

Figure\,3 shows, traced over the same CMD of Figure\,2, the MS ridge
line (solid curve) obtained by applying a sigma-clipping method to the
colour of the stars along the MS, as explained in Section\,4. The
dashed line in the same figure corresponds to the theoretical isochrone
computed by Baraffe et al. (1997) for a 10\,Gyr old cluster with
metallicity $[M/H]=-1$ and Helium content $Y=0.25$, as is appropriate
for NGC\,6218. The excellent fit shown in Figure\,3 corresponds to a
distance modulus  $(m-M)_{\rm V}=14.23$ and colour excess
$E(B-V)=0.18$, which compare favourably with the recently obtained
$(m-M)_{\rm V}=14.11$ and colour excess $E(B-V)=0.19$ of Hargis \&
Sandquist (2004). The difference in the distance modulus is small and
fully consistent with the uncertainty on our absolute photometric
calibration. We note, however, that the difference would vanish if we
were to take the distance modulus of Sato et al. (1989), namely
$(m-M)_{\rm V}=14.25$. We note here that the small discrepancy
between the observed MS ridge line and that predicted by theoretical
models at the lowest mass end (see Figure\,3) is a well known
limitation of the theory and most probably stems from the lack of a
proper treatment of the TiO molecular opacity, as extensively discussed
in Baraffe et al. (1998; see also Chabrier 2001). The discrepancy
becomes progressively smaller at longer wavelengths (Pulone et al.
2003; Pulone et al. 1998). In any case, as we explain in the following
section, this small uncertainty does not affect our conclusions on the
mass distribution of MS stars, because over the mass range covered by
our observations ($\sim 0.3 - 0.8$\,\Msolar) an uncertainty of
$0.1$\,mag in $V$ translates to an uncertainty of $\sim 0.01$\,\Msolar
on the mass.   

In order to study possible changes of the stellar population with
position in the cluster, we have defined four distinct concentric 
regions over the area of our images. The innermost three are annuli
$30\arcsec$ wide while the fourth region includes all objects farther
than $90\arcsec$ from the cluster centre through to the edge of the
frame. For simplicity, hereafter we refer to these regions as "centre"
and "rings" 1 through 3. They contain, respectively, 1664, 4234, 5469
and 4648 objects. If instead of fixing the width of the rings we had
kept constant their area or the number of stars in them, the radial
extent of the annuli would have obviously changed. On the other hand,
we have verified that the progressive variation of the CMD and LF with
radius is largely insensitive to the exact size of the rings. 

\begin{figure*}
\centering
\includegraphics[width=16cm]{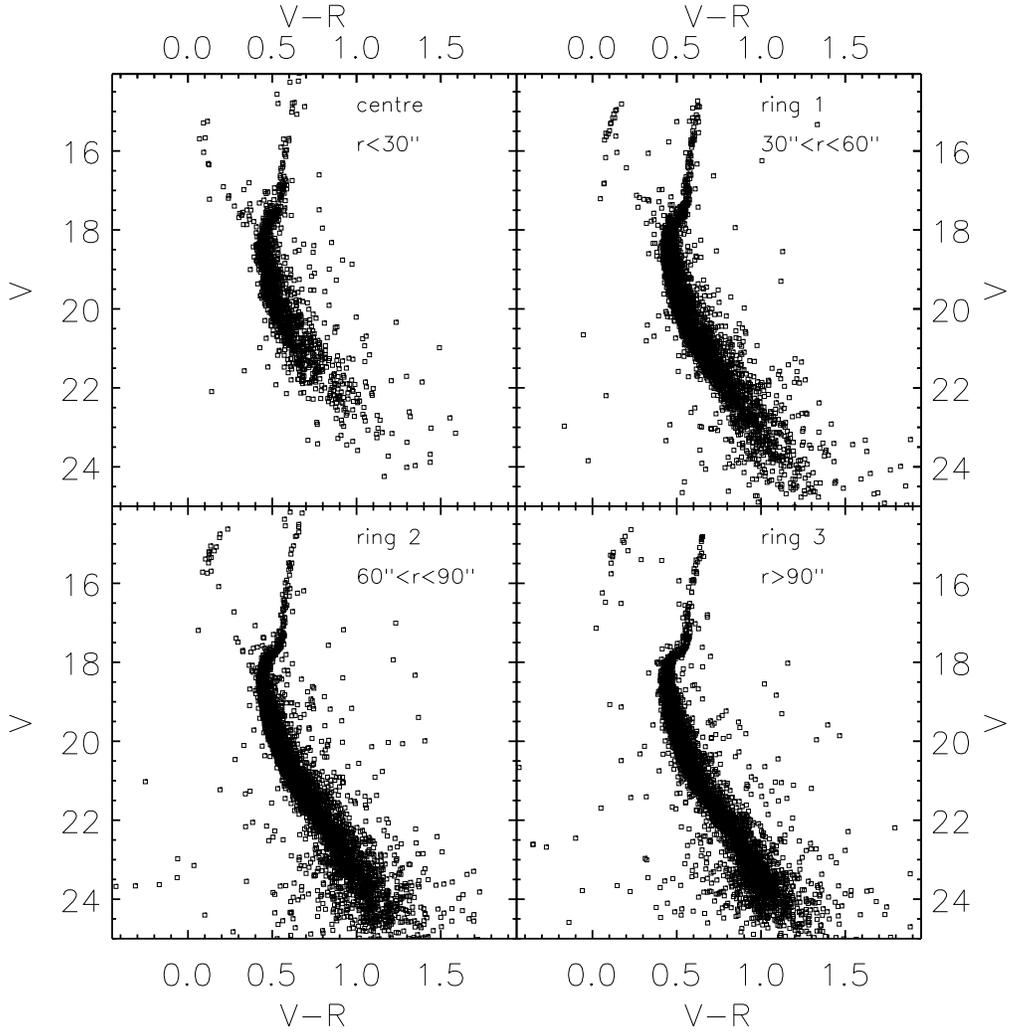} 
\caption{Individual CMDs of the four regions in which we have divided
our NGC\,6218 field.} 
\label{fig4}
\end{figure*}

The CMDs of the individual regions are shown in the four panels of 
Figure\,\ref{fig4}. The most notable differences among the four panels
is the marked central concentration of the blue stragglers, which are
totally absent in the third ring, and the remarkable lack of bright red
giants in the cluster centre, where the red giant branch stops one to
two magnitudes fainter than outside. These issues are discussed in a
companion paper (Sabbi et al. 2005). 

Also noteworthy in Figure\,\ref{fig4} is the rapid fading of the MS in
the cluster centre well before the onset of significant photometric 
incompleteness. For instance, at $V\simeq 22$ the completeness is still
above 50\,\% (see below) and the photometric error small ($\sigma_{\rm
V}=0.03$, $\sigma_{\rm R}=0.03$), but the MS is already so sparse in
the CMD that it cannot be distinguished from field stars. This finding
is already a clear indication of the mass segregation that accompanies
the normal dynamical relaxation process of GCs. Due to the
equipartition of energy among stars, less massive objects tend to
increase their kinetic energy and move onto larger orbits that force
them outwards, away from the cluster centre, for most of their orbital
period.

\section{The luminosity and mass function} 

In order to study this effect in a more quantitative way, we decided to
derive the LF of MS stars at various locations
inside the cluster. In particular, one has to pay special attention to
photometric incompleteness, which acts differentially on the LF and
affects more prominently the denser central regions and less or
marginally the outer rings. Incompleteness is due to crowding and to
saturated stars, whose bright halo can mask possible faint objects in
their vicinity, both more likely to affect substantially the central
regions (see Figure\,\ref{fig1}). If ignored or corrected for in a
uniform way across the whole image, by applying the same correction to
all regions, these effects would bias the final results and mimic the
presence of a high degree of mass segregation. For this reason, we
conducted a series of artificial star experiments for the four
individual regions (core and three rings), using the most appropriate
local PSF for each. 

\begin{table} 
\centering
\caption{Photometric completeness $f$ and the associated
$1\,\sigma$ uncertainty, both in percent, in the four regions in which 
we have divided our NGC\,6218 field.} 
\begin{tabular}{ccccccccc}
\hline 
  &  \multicolumn{2}{c}{\em Centre} & \multicolumn{2}{c}{\em Ring 1}& 
     \multicolumn{2}{c}{\em Ring 2} & \multicolumn{2}{c}{\em Ring 3}\\ 
{\em $ V $} & 
  $f$ & $\sigma$ & $f$ & $\sigma$ & $f$ & $\sigma$ & $f$ & $\sigma$ \\
\hline
 18.25 & 91 & 2 & 94 & 1 & 95 & 2 & 96 & 1 \\
 18.75 & 86 & 3 & 93 & 2 & 94 & 2 & 96 & 1 \\
 19.25 & 79 & 6 & 92 & 4 & 93 & 2 & 95 & 1 \\
 19.75 & 76 & 5 & 90 & 2 & 91 & 1 & 94 & 1 \\
 20.25 & 75 & 4 & 86 & 4 & 89 & 1 & 92 & 1 \\
 20.75 & 67 & 7 & 82 & 2 & 86 & 2 & 90 & 1 \\
 21.25 & 59 & 9 & 77 & 2 & 83 & 2 & 88 & 2 \\
 21.75 & 47 & 6 & 70 & 2 & 80 & 2 & 84 & 2 \\
 22.25 & 34 & 5 & 62 & 5 & 74 & 2 & 80 & 2 \\
 22.75 & 22 & 4 & 48 & 4 & 63 & 2 & 72 & 1 \\
 23.25 & 13 & 4 & 31 & 3 & 49 & 3 & 59 & 2 \\
 23.75 &  6 & 3 & 18 & 3 & 33 & 2 & 42 & 2 \\
 24.25 &  2 & 1 &  7 & 3 & 19 & 4 & 23 & 2 \\ 
\hline
\end{tabular}
\vspace{0.5cm}
\label{tab2}
\end{table}

The artificial star tests were run on the combined images, in both
bands. For each $0.5$ magnitude bin we carried out 10 trials by adding
a fraction of 10\,\% of the total number of objects (see Section 2). 
These trials were followed by running the tasks {\it daophot.daofind},
{\it daophot.allstar} and {\it daophot.allframe}, with the same
parameters used in the reduction of the scientific images so that we
could assess the fraction of objects recovered by the procedure and the
associated photometric errors. The resulting photometric completeness
is given in Table\,2 for each region as a function of the $V$
magnitude, along with the associated $1\,\sigma$ uncertainty.

In order to derive the LF from the CMDs of Figure\,\ref{fig4}, one
needs to separate true MS stars from field objects. The latter are
however not significant at the Galactic latitude of NGC\,6218
($b=26.3$), as Figure\,\ref{fig2} already shows. The Galaxy model of
Ratnatunga \& Bahcall (1985) predicts about 17 field stars per
arcmin$^2$ towards the direction of NGC\,6218 down to magnitude $V=25$,
with about half of them in the range $23 < V < 25$. This would
correspond to $\sim 200$ contaminating field stars in our FOV down to
$V=25$ of which  $\sim 100$ brighter than $V=23$. The CMD of
Figure\,\ref{fig2} has of order 2,000 stars in the range $V\ge 23$,
before any correction for incompleteness, and there are approximately
14,000 stars brighter than that magnitude. Therefore, it would seem
perfectly unnecessary to take field star contamination into account. 

On the other hand, since field star contamination affects more
prominently the lower end of the MS, which is most relevant to our
investigation, we decided to remove it at all magnitudes. To this aim,
we made use of the colour information in the CMD and applied the
$\sigma$  clipping criterion described by De Marchi \& Paresce (1995)
to identify the possible outliers. In practice, from the CMDs of
Figure\,\ref{fig4} we measured the LF of MS stars by counting the
objects in each 0.5 mag bin and within $\pm 5$ times the colour
standard deviation around the MS ridge line and rejected the rest as
field objects. This procedure identifies about $350$ field stars in the
magnitude range $18 < V < 23$ and about $50$ at fainter magnitudes.
When photometric incompleteness is taken into account, the resulting
number of contaminating stars is approximately a factor of four higher
than that predicted by the model of Bahcall \& Soneira (1985). It is,
nevertheless, very small ($\sim 2\,\%$) when compared to the rest of
bona fide MS stars and, in each magnitude bin, is smaller than the
statistical uncertainty associated with the counting process. In the
following, we will consider only bona fide MS stars as defined in this
way but, as will become clear in the rest of the paper, our conclusions
would remain unchanged had we decided to ignore field star
contamination and treat all stars as cluster members. 


\begin{table} 
\centering
\caption{Average main sequence fiducial points and colour width}
\begin{tabular}{ccc}
\hline 
 $V$ & $V-R$ & $\sigma_{\rm V-R}$ \\ 
\hline
 18.25 &   0.447   &   0.016 \\
 18.75 &   0.450   &   0.017 \\
 19.25 &   0.473   &   0.021 \\
 19.75 &   0.505   &   0.027 \\
 20.25 &   0.547   &   0.029 \\
 20.75 &   0.603   &   0.035 \\
 21.25 &   0.670   &   0.045 \\
 21.75 &   0.750   &   0.050 \\
 22.25 &   0.840   &   0.058 \\
 22.75 &   0.915   &   0.064 \\
 23.25 &   0.971   &   0.078 \\ 
 23.75 &   1.034   &   0.094 \\ 
 24.25 &   1.091   &   0.131 \\ 
 24.75 &   1.173   &   0.150 \\ 
\hline
\end{tabular}
\vspace{0.5cm}
\label{tab3}
\end{table}

\begin{table*} 
\centering
\caption{Luminosity functions measured in each of the four regions in
which we have divided our NGC\,6218 field. For each region, the table
gives as a function of the $V$-band magnitude the number
of stars per half-magnitude bin before ($N_{\rm o}$) and after ($N$)
completeness correction and the uncertainty on $N$.}
\begin{tabular}{ccccccccccccccccc}
\hline 
     & & \multicolumn{3}{c}{\em Centre} & & \multicolumn{3}{c}{\em Ring 1}  
     & & \multicolumn{3}{c}{\em Ring 2} & & \multicolumn{3}{c}{\em Ring 3}\\ 
 $V$ & & $N_{\rm o}$ & $N$ & $\sigma_{\rm N}$  
     & &   $N_{\rm o}$ & $N$ & $\sigma_{\rm N}$ 
     & &   $N_{\rm o}$ & $N$ & $\sigma_{\rm N}$ 
     & &   $N_{\rm o}$ & $N$ & $\sigma_{\rm N}$ \\
\hline
18.25 && 182 & 200 & 15 && 335 & 356 & 20 && 345 & 363 & 21 && 228 & 238 & 16 \\
18.75 && 199 & 231 & 18 && 396 & 426 & 23 && 412 & 438 & 24 && 312 & 325 & 18 \\
19.25 && 224 & 284 & 29 && 474 & 515 & 33 && 509 & 547 & 27 && 380 & 400 & 21 \\
19.75 && 216 & 284 & 27 && 530 & 589 & 29 && 534 & 587 & 26 && 448 & 477 & 23 \\
20.25 && 181 & 241 & 22 && 469 & 545 & 36 && 557 & 626 & 27 && 449 & 488 & 24 \\
20.75 && 139 & 207 & 28 && 401 & 489 & 27 && 507 & 590 & 30 && 377 & 419 & 22 \\
21.25 && 104 & 176 & 32 && 320 & 416 & 26 && 436 & 525 & 28 && 322 & 366 & 22 \\
21.75 &&  70 & 149 & 26 && 257 & 367 & 25 && 358 & 448 & 26 && 312 & 371 & 23 \\
22.25 &&  46 & 135 & 28 && 154 & 248 & 28 && 307 & 415 & 26 && 291 & 364 & 23 \\
22.75 &&  21 &  95 & 27 && 166 & 346 & 39 && 250 & 397 & 28 && 285 & 396 & 24 \\
23.25 &&  15 & 115 & 46 && 102 & 329 & 46 && 268 & 547 & 47 && 330 & 559 & 36 \\
23.75 &&   5 &  83 & 56 &&  94 & 522 &102 && 239 & 724 & 64 && 297 & 707 & 53 \\
24.25 &&   2 & 100 & 87 &&  45 & 643 &292 && 177 & 932 &208 && 190 & 826 & 94 \\
\hline
\end{tabular}
\vspace{0.5cm}
\label{tab4}
\end{table*}

The average MS fiducial points are drawn as a solid line in Figure\,3
and are listed in Table\,3 together with the colour width of the MS
($\sigma_{\rm V-R}$). Table\,4 gives the four LFs, before and after
correction for photometric incompleteness, and the corresponding rms
errors coming from the Poisson statistics of the counting process (only
for the LF corrected for incompleteness). All values have been rounded
off to the nearest integer. The data are also shown graphically in
Figure\,5, where the squares give the LF of the four regions, with
the centre at the bottom and the outermost ring at the top. As
Figure\,5 shows, all four LFs are remarkably flat, with the two central
rings displaying a sensible drop in the number counts at low masses for
$M_V>6.5$ (see upper axis). This result is very robust, because only
data-points with an associated photometric completeness in excess of
50\,\% are shown in Figure\,5. 

\begin{figure}
\centering
\resizebox{\hsize}{!}{\includegraphics[width=16cm]{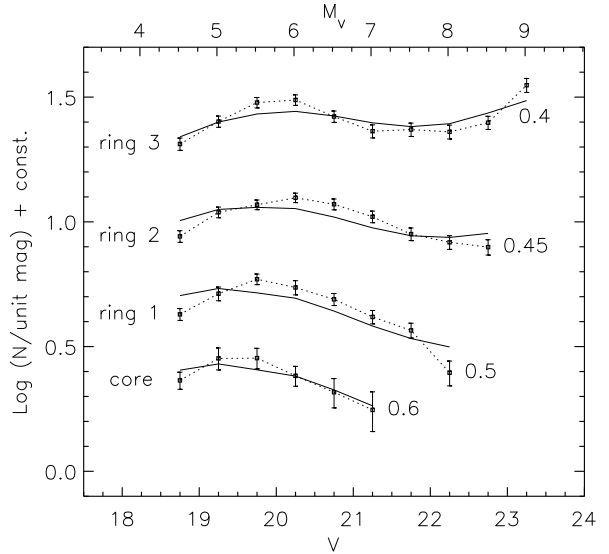}}
\caption{The squares are the LF of the four regions in NGC\,6218, after
correction for photometric incompleteness, as given in Table\,4. The
model LFs that best fit the data are shown as solid lines. The index of
their corresponding power-law MF is, from bottom to top, $\alpha=1.4, 
1.3, 0.6$ and $0.1$. A positive index means that the number of stars
decreases with mass. The MS stellar mass corresponding to the faintest
data point of each LF is indicated, in units of \Msolar.}
\label{fig5}
\end{figure} 

A flat or dropping LF (i.e. one that drops with increasing magnitude)
is not unusual in the core of GCs, where the effects of mass
segregation are strongest. Such is, for instance, the case of 47\,Tuc
(Paresce, De Marchi \& Jedrzejewski 1995), NGC\,6397 (King, Sosin \&
Cool 1995) and NGC\,7078 (De Marchi \& Paresce 1996; Pasquali et al.
2004). What is most unusual, however, is a flat LF near the cluster's
half-mass radius, where a sample of 12 GCs studied by us (Paresce \& De
Marchi 2000) has revealed a remarkably consistent behaviour with a LF
increasing monotonically from the TO luminosity to $M_V\simeq 10$. For
NGC\,6218 this is clearly not the case, since the top LF in Figure\,5
samples the stellar population from a radius of $90\arcsec$ through to
$155\arcsec$, with an equivalent radius of $115\arcsec$, comfortably
reaching the nominal cluster's half-mass radius at $r_{\rm
h}=2\farcm16$ (Harris 1996). While near the half-mass radius of
NGC\,6397 the number of stars per unit magnitude grows by a factor of
about four from the TO luminosity  to $M_V \simeq 10$ (King et al.
1998), this number remains practically constant for NGC\,6218. So far,
the only GCs known to have a dropping or flat LF near the half-mass
radius are NGC\,6712 (De Marchi et al. 1999) and Pal\,5 (Koch et al.
2004) and both are expected to have undergone a strong dynamical
interaction with the Galactic tidal field, at variance with the
expectations for NGC\,6218. 

The LFs shown in Figure\,5 can in principle be converted into MFs,
since we have shown in Figure\,3 that the cluster MS is well fitted,
within the theoretical and observational errors, by the  available
theoretical models of low-mass stars (Baraffe et al. 1997). We can,
thus, reasonably expect these models to provide a reliable relationship
between luminosity and mass. On the other hand, in order to keep
observational errors clearly separated from theoretical uncertainties,
we prefer to fold a model MF through the derivative of the
mass--luminosity (M--L) relationship and compare the resulting model LF
with the observations. The solid lines in Figure\,5 are the theoretical
LFs obtained by multiplying a simple power-law MF of the type $dM/dm
\propto m^\alpha$ by the derivative of the M--L relationship. Although
the MF of GCs is more complex than a simple power-law, particularly at
low masses (Paresce \& De Marchi 2000), over the narrow mass range
($0.4 - 0.8$\,\Msolar) spanned by these observations this simplifying
assumption is valid (De  Marchi, Paresce \& Portegies Zwart 2005), . 

With the adopted M--L relationship, a distance modulus $(m-M)_{\rm
V}=14.23$ and colour excess $(E(B-V)=0.18$ (see Section\,3), $V=18.8$
corresponds to the TO mass of $0.8$\,\Msolar. The lowest mass reached
in each region is indicated in Figure\,5 and varies from $\sim
0.6$\,\Msolar in the core to $\sim 0.4$\,\Msolar in ring 3. The 
power-law MF indices that best fit the data are indicated in the figure
caption and range from $\alpha = 1.4$ in the core to $\alpha =  0.1$ in
the outermost ring near the half-mass radius. The fact that these
values are all positive implies that the number of stars is decreasing
with mass. With the notation used here, the canonical Salpeter IMF
would have $\alpha=-2.35$. For comparison, over the same mass range
spanned by the MF of ring 3, the power-law index that best fits the MF
of NGC\,6397 is $\alpha \simeq -1.6$ and that of the other 11 GCs in
the sample studied by Paresce \& De Marchi (2000) is of the same order.
It is, therefore, clear that NGC\,6218 is surprisingly devoid of
low-mass stars, at least out to its half-mass radius.

\section{Dynamical structure of NGC\,6218}

In order to clarify the origin of this deficiency of low-mass stars, it
is useful to compare the radial MF variation implied by Figure\,5 with
that expected from the two-body relaxation process that operates in any
stellar cluster. To study the dynamical properties of the cluster, we
employed the multi-mass Michie--King models originally developed by
Meylan (1987; 1988) and later suitably modified by Pulone et al. (1999)
and De Marchi et al. (2000) for the general case of clusters with a set
of radially varying LFs. Our parametric modeling approach assumes
energy equipartition among stars of different masses. Each model run is
characterised by a MF in the form of a power-law with a variable index
$\alpha$, and by four structural parameters describing, respectively,
the scale radius ($ r_{\rm c}$), the scale velocity ($v_{\rm s}$), the
central value of the dimensionless gravitational potential $W_{\rm o}$,
and the anisotropy radius ($r_{\rm a}$). 

From the parameter space defined in this way, we selected those models
that simultaneously fit both the observed surface brightness profile
(SBP) and the central value of the velocity dispersion as measured,
respectively, by Trager et al. (1995) and by Pryor et al. (1988).
However, while forcing a good fit to these observables constrains the
values of $r_{\rm c}$, $v_{\rm s}$, $W_{\rm o}$, and $r_{\rm a}$, the
MF can still take on a variety of shapes. To break this degeneracy, we
imposed the additional condition that the model LF agrees with that
observed at all available locations simultaneously. This, in turn, sets
very stringent constraints on the PGMF, i.e. on the MF of the cluster
as a whole. 

For practical purposes, the model PGMF has been divided into sixteen
different mass classes, covering main sequence stars, white dwarfs and
heavy remnants, precisely as described in Pulone et al. (1999). We ran
a large number of trials looking for a suitable shape of the PGMF such
that the local MFs implied by mass segregation would locally fit the
observations. As explained in the previous section, in order to keep
observational errors and theoretical uncertainties separate, we
converted the model MFs into to LFs, using the same M--L relation,
distance modulus and colour excess of Figure\,5, and compared those to
the observations.

\begin{figure}
\centering
\resizebox{\hsize}{!}{\includegraphics[width=16cm]{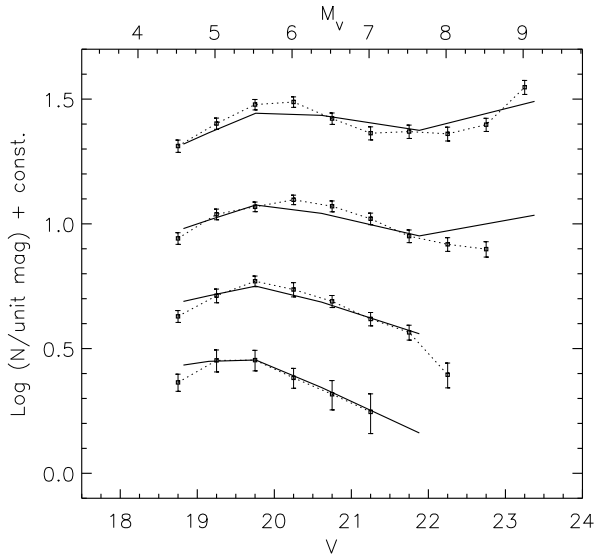}}
\caption{Same LFs as those of Figure\,5, but here the solid lines show
the LFs predicted by our multi-mass Michie--King model at various radii
inside the cluster, starting from the same PGMF with index $\alpha =
0.1$.  The expected radial variation of the LF is fully consistent with
the observations.}
\label{fig6}
\end{figure}

The set of model LFs that best fits all available observations is shown
in Figure\,6 and is drawn from the same PGMF with index $\alpha=0.1$
for stars less massive than $0.8$\,\Msolar. Figure\,7 illustrates the
remarkably good fit to the SBP, whereas the values of the best fitting
structural parameters are given in Table\,5, where they can be compared
with those in the literature. The agreement is excellent, apart from a
small difference in the value of the tidal radius which is, admittedly,
not seriously constrained by our data. We note here that we can
directly compare the observed SBP with our model since the solid line
in Figure\,7 corresponds to stars of $\sim 0.8$\,M$_\odot$, which are
those contributing most of the cluster's light. As one should expect,
stars in different mass classes have different projected radial
distributions, with the relative density at any location governed by
the relaxation process.

\begin{table}[b]
\caption[]{Cluster structural parameters for NGC\,6218}
\begin{tabular}{lccc}
\hline
Parameter & Fitted & Literature & Ref.\\ & value & value & \\
\hline
core radius $r_{\rm c}$   & $0\farcm6$  & $0\farcm7$   & $a$\\
tidal radius $r_{\rm t}$  & $17\farcm6$    & $20\farcm4$ & $a$\\
half-mass radius $r_{\rm h}$ & $3\farcm0$ & $2\farcm2$& $a$\\
central vel. disp. $\sigma_{\rm v}$ & 4.5 km~s$^{-1}$ & 4.0 km~s$^{-1}$
& $b$\\
\hline
\end{tabular}
\par\noindent
$a$: Harris (1996)
\par\noindent
$b$: Pryor \& Meylan (1993)
\label{tab5}
\end{table}

\begin{figure}[t]
\centering
\resizebox{\hsize}{!}{\includegraphics[width=16cm]{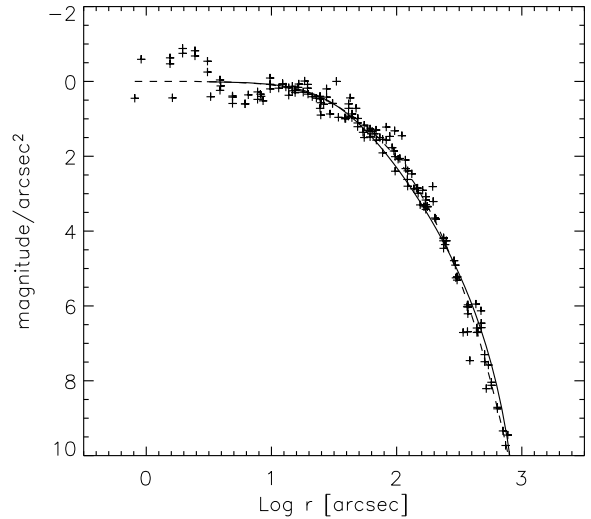}}
\caption{The surface brightness profile of NGC\,6218 (crosses, from
Trager et al. 1995) is well reproduced by our dynamical model (solid
line). The observations are normalised to the central value of the best
fitting profile (dashed line) as given by Trager et al. (1995).}
\label{fig7}
\end{figure}

As for the shape of the PGMF, $\alpha=0.1$ implies that it is rather 
flat or even decreasing with mass. This may have already been inferred
by the flat shape of the MF measured near the half-mass radius $r_{\rm
h}$, as it is  known that the latter is a good approximation to the
cluster's PGMF (Richer et al. 1991, De Marchi et al. 1995).
Nevertheless, it is comforting to see that a PGMF of this type is
consistent with the shape of the MF observed elsewhere in the cluster,
suggesting that the radial change of the MF with radius is the result
of the relaxation process and that, therefore, the cluster is in
dynamical equilibrium, at least out to the half-mass radius.

This also allows us to set some constraints on the shape of the MF for
stars above the TO mass, namely those that have now evolved into
degenerate objects (white dwarfs, neutron stars or even black holes).
Although not as tightly constrained as for MS stars, the power-law
index $\alpha$ of stars more massive than the TO is much closer to the
Salpeter value, with $\alpha=-2$ giving the best fit. This parameter
determines the fraction of heavy remnants in the cluster, which we find
to be of order $60\,\%$, and affects the overall distribution of the
stars in all other classes of mass, to which the SBP is rather
sensitive. We find that values of $\alpha$ larger or smaller than 2
give a progressively worse fit to the SBP, which becomes unacceptable
for $\alpha > -1.5$ or $\alpha < -2.5$.   

The central velocity dispersion predicted by our model (see Table\,5)
agrees well with the observed value reported by Pryor \& Meylan (1993)
and implies  a total cluster mass of $\sim 1.2 \times 10^5$\,\Msolar,
also in line with that of $\sim 1.3 \times 10^5$\,\Msolar derived by
those authors. The mass to light ($M/L$) ratio that we obtain depends
on the assumed total luminosity of the cluster, which appears to be
rather uncertain in the literature. The total $V$-band apparent
magnitude varies from $V=6.05$ of Peterson \& Reed (1987; also adopted
by Djorgovski 1993) to $V=6.77$ of Webbink (1985). In his recent
catalogue, Harris (1996; revision 2003) gives $V=6.70$, corresponding
to a total absolute magnitude $M_{\rm V} = -7.31$ and a total
luminosity of $\sim 7.2 \times 10^4$\,L$_{\odot}$. With the latter
value, we obtain $M/L_{\rm V}=1.7$ for the whole cluster and $M/L_{\rm
V}=1.6$ inside the core radius.

\section{Discussion and conclusions}

What does a flat PGMF tell us about NGC\,6218? Using a sample of 12
halo GCs for which deep HST observations are available, Paresce \& De
Marchi (2000) showed that their PGMF must directly reflect the
properties of the IMF. This conclusion is based on the observation that
those clusters have very different properties (total mass, metallicity,
concentration and space  motion parameters) but they show, in practice,
the same or very similar PGMF. This would be hard to justify if the
clusters were initially born with very different IMFs. The underlying
common IMF, exemplified by that of NGC\,6397, is best represented by a
power-law distribution that tapers off below $\sim 0.3$\,\Msolar (see
Paresce \& De Marchi 2000 and De Marchi et al. 2005 for details). 

The PGMF of NGC\,6218 clearly does not match that of the 12 objects in
the sample of Paresce \& De Marchi (2000), because it is remarkably
flat. However, at least two other cases of flat or even dropping PGMF
have been reported, namely those of NGC\,6712 (De Marchi et al. 1999)
and Pal\,5 (Koch et al. 2004), with NGC\,6712 revealing an even 
stronger deficit of low-mass stars than NGC\,6218. On the other hand,
there are good observational reasons to believe that both NGC\,6712 and
Pal\,5 have suffered severe tidal disruption  that has considerably
altered their original distribution of stellar  masses. For instance,
Pal\,5 has well defined tidal tails extending over $10^\circ$ across
the sky (Odenkirchen et al. 2001). But, more generally, the present
total mass and space motion parameters of these clusters imply that
they have some of the highest destruction rates in the whole Galactic
GC system. Gnedin \& Ostriker (1997) predict a remaining lifetime as
low as $\sim 0.3$\,Gyr for NGC\,6712 and $\sim 1$\,Gyr for Pal\,5,
whereas Dinescu et al. (1999) give a time to disruption of $\sim
3.9$\,Gyr for NGC\,6712 and just $\sim 0.1$\,Gyr for Pal\,5,
respectively. Although not in agreement with one another, these sets of
values are far lower than the average time to disruption, which for
both authors is of order 12\,Gyr (and thus comparable with the typical
GC age). 



Quite surprisingly, however, the total mass and space motion parameters
of NGC\,6218 do not seem to put this cluster in any imminent danger. In
fact, the estimated time to disruption for this object varies from
$\sim 23.5$\,Gyr (Gnedin \& Ostriker 1997) to $\sim 33$\,Gyr (Dinescu
et al. 1999). Similarly, collisional N-body simulations of a cluster
with the properties of NGC\,6218 moving in an external tidal field
produce a total lifetime for this object of $\sim 29$\,Gyr (H.
Baumgardt, private comm.; see Baumgardt \& Makino 2003 for model
details). Assuming a GC age of $\sim 12.5$\,Gyr (Krauss \& Chaboyer
2003), this implies $T_{\rm d} \simeq 16.5$\,Gyr for NGC\,6218. In
other words, none of these models implies a dynamically troubled past
for NGC\,6218, leaving in principle open the possibility that it was
born with a rather flat IMF, at least in the mass range covered by our
observations. While such a hypothesis cannot {\em a priori} be
excluded, this would be the first case known. 

On the other hand, models of the dynamical interaction of GCs with the
Galactic tidal field are, unfortunately, still subject to large
uncertainties. As both Gnedin \& Ostriker (1997) and Dinescu et al.
(1999) point out, different assumptions on the initial conditions
(cluster orbits) and  on the Galactic potential can result in rather
different destruction rates for the same cluster. While these models
are useful to address the replenishment of the halo over time from the
disruption of individual clusters, and therefore helpful to understand
the global properties of the GC system and its evolution, they may be
still too crude to precisely describe the past history of individual
clusters. 

We, however, believe that the apparent discrepancy between the
predicted value of $T_{\rm d}$ and the shape of the PGMF of NGC\,6218
is simply due to the wrong assumption as to the cluster orbit. In fact,
the space motion parameters used by these authors for NGC\,6218 are not
consistent with the latest determination of its absolute proper motion
based on the Hipparcos reference system. The  proper motion measured by
Brosche et al. (1991), combined with the radial velocity measurements
of Pryor \& Meylan 1993, allowed Dauphole et al. (1996) to put some
constraints on the space motion parameters of this cluster. Their
findings were confirmed by an independent analysis of the orbit, based
on improved absolute proper motions (Scholz et al. 1996), which
indicated that NGC\,6218 should have a short orbital period 
($0.17$\,Gyr) but also that it never ventures closer than $\sim 3$\,Kpc
from the Galactic centre, with less than 15\,\% of its orbit lying
within 1\,Kpc of the Galactic plane. However, a more recent study of
the orbit of NGC\,6218, based on the new Hipparcos reference system
(ESA 1997), has led Odenkirchen et al. (1997) to the conclusion that
NGC\,6218 has a highly irregular motion. In particular, they find that
the low value of the cluster's axial angular momentum forces it to pass
the Galactic centre at short distance ($R_{\rm p} = 0.6$\,kpc) and  to
get into strong interaction with the Galactic bulge, since the bulge
destruction rate scales with the fourth power of $R_{\rm p}$ (Dinescu
et al. 1999).

With an orbit of this type, the total lifetime of NGC\,6218 predicted
by the models of Baumgardt \& Makino (2003; H. Baumgardt priv. comm.)
would decrease from 29\,Gyr to 17\,Gyr, thus implying a time until
disruption of only $T_{\rm d} \simeq 4.5$\,Gyr, assuming a typical GC
age of $\sim 12.5$\,Gyr as above (Krauss \& Chaboyer 2003). Although
not as small as that of NGC\,6712 or Pal\,5, this value of $T_{\rm d}$
places NGC\,6218 among the clusters at higher risk of disruption and
suggests that a considerable fraction of its original stellar
population should have been stripped from the cluster. 

We can estimate the amount of mass lost by NGC\,6218 in the hypothesis
mentioned above that all GCs were born with a very similar IMF (Paresce
\& De Marchi 2000). As for the latter, we use the tapered power-law
proposed by De Marchi et al. (2005) and find that the present total
mass due to MS stars is about one fifth of the original $\sim 6 \times
10^5$\,\Msolar. This value is in full agreement with the revised
calculations of H. Baumgardt (priv. comm.) suggesting an initial mass
of $7.6 \times 10^5$\,\Msolar. A natural question to pose is when this
major mass loss process took place.  

If the cluster is in thermo-dynamical equilibrium, one only needs
to look at the half-mass relaxation time in order to answer this
question. The data are compatible with this hypothesis, since our model
imposes the condition of energy equipartition and Figure\,6 and
Table\,5 show that this is consistent with the radial variation of the
MF and the cluster's structural parameters.\footnote{We are grateful to
an anonymous Referee for pointing out that, although plausible,
thermo-dynamical equilibrium is not necessarily implied by the data. In
fact, the good fit shown in Figure\,6 and Table\,5 is a necessary
condition for energy equipartition, but it is not a sufficient one. In
principle, there could be models that fit the data equally well but in
which the equipartition of energy is not complete. Unfortunately, the
method that we followed (Meylan 1987; 1988) does not allow us to
investigate this hypothesis.} Under this assumption, the half-mass
relaxation time suggested by our models is  $t_{\rm rh}\simeq
0.5$\,Gyr. This value is slightly lower than the $t_{\rm rh} \simeq
0.7$\,Gyr given by Djorgovski (1993) and $t_{\rm rh}  \simeq 1$\,Gyr of
Gnedin \& Ostriker (1997), although still compatible with them, because
the MF assumed by these authors is very steep at the low mass end,
contrary to what we find, thus resulting in a much larger number of
objects in the cluster. In any case, it seems possible to exclude that
a major mass loss episode happened in the course of the past 1\,Gyr or
so, since in that case the cluster should have not yet reached a
condition of equilibrium. The fact that tidal tails are very tenuous
around NGC\,6218 (if at all present; see Lehmann \& Scholz 1997) gives
support to this scenario. Mass loss must, therefore, have happened
either long ago or continuously, over the cluster lifetime, at a low
rate of about $5 \times 10^3$\,\Msolar per orbit. 

The present data do not allow us to explore the past dynamical history
of NGC\,6218 in more detail. Nevertheless, our results show the
importance of feeding models of GC distruption with reliable cluster
orbits, since the strength and extent of the interaction between the
Galaxy and GCs can vary dramatically with the orbit. However, it also
depends on the gravitational potential of the Galaxy, and particularly
on that of the  bulge, thereby making it crucial for models of this
type to rely on a solid observational description of the structure of
the Galaxy and of its components (thin/thick disc, bulge, halo), which
is presently lacking. This information should become available in the
near future with the advent of missions such as SIM and Gaia. But
already now, if the cluster orbit is reasonably well understood,
observations of the cluster PGMF can set meaningful constraints on the
mass distribution in the Galaxy. Indeed, the PGMF is a more reliable 
indicator of the past dynamical history of a cluster than, for
instance, its present location and space motion parameters or even the
presence and extent of its tidal tails. The position and velocity of a
cluster are instantaneous quantities and can vary largely in time.
Tidal tails are made up of unbound stars and, as such, are short lived
and can only probe the immediate past of a cluster. The PGMF, on the
other hand, reflects the integrated effect of the interaction with the
Galaxy throughout the whole life of the cluster and provides an
indication of the amount of mass lost to the Galaxy. By measuring the
PGMF of a sizeable number of clusters, including those with chaotic or
anyhow irregular orbits, it should be possible to set meaningful
constraints on the form of the Galactic potential well before the
availability of astrometric data from interferometric space
observatories.

\begin{acknowledgements}

It is a pleasure to thank Dana Dinescu and Holger Baumgardt for very
useful discussions and the latter also for providing us with the
results of his simulations ahead of publication. 

\end{acknowledgements}

\end{document}